FERMILAB-CONF-16-506-AD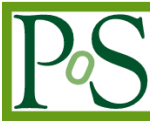
PoS — PROCEEDINGS OF SCIENCE# Fermilab's Accelerator Complex: Current Status, Upgrades and Outlook

**M. E. Convery**[1]

*Fermi National Accelerator Laboratory*
*Batavia, IL 60510, USA*
*E-mail:* `convery@fnal.gov`We report on the status of the Fermilab accelerator complex, including recent performance, upgrades in progress, and plans for the future. Beam delivery to the neutrino experiments surpassed our goals for the past year. The Proton Improvement Plan is well underway with successful 15 Hz beam operation. Beam power of 700 kW to the NOvA experiment was demonstrated and will be routine in the next year. We are also preparing the Muon Campus to commission beam to the g-2 experiment.*38th International Conference on High Energy Physics*
*3-10 August 2016*
*Chicago, USA*---

[1]Speaker

Operated by Fermi Research Alliance, LLC under Contract No. DE-AC02-07CH11359 with the United States Department of Energy.ⓒ Copyright owned by the author(s) under the terms of the Creative Commons
Attribution-NonCommercial-NoDerivatives 4.0 International License (CC BY-NC-ND 4.0).     http://pos.sissa.it/Operated by Fermi Research Alliance, LLC under Contract No. DE-AC02-07CH11359 with the United States Department of Energy

1.     **Introduction**

The Fermilab accelerator complex, shown in Fig. 1, has surpassed its beam delivery goals for the past year, has demonstrated 700 kW beam power to the NOvA experiment, and is preparing the Muon Campus to begin commissioning beam to the g-2 experiment. These accomplishments would not be possible without the successful execution of the Proton Improvement Plan (PIP) which, among many other improvements, has enabled beam from the Booster at 15 Hz. Other near-term upgrades are planned for increasing beam flux to the Short Baseline Neutrino experiments. For the longer term, the PIP-II project is underway to provide protons with more than 1 MW beam power to the Long Baseline Neutrino Facility.

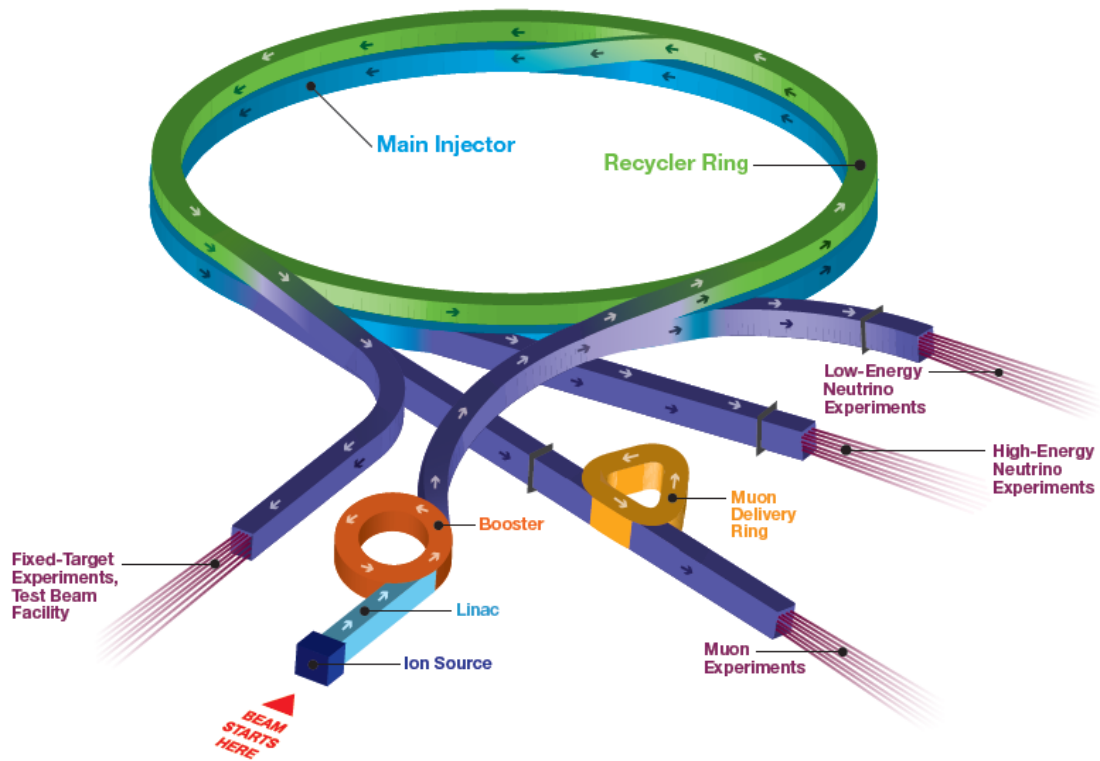

Figure 1: The Fermilab accelerator complex.

2.     **Beam power of 700 kW to the NuMI target**

Prior to the Accelerator NOvA Upgrade (ANU) in 2013, 8-GeV protons from the Booster were RF slip-stacked in the Main Injector, ramped to 120 GeV, and extracted to the NuMI target every 2.2s, for a beam power of about 320 kW. The Recycler, a ring with fixed 8 GeV kinetic energy which shares a tunnel with the Main Injector, was used for storing antiprotons for the Tevatron collider. ANU provided RF in the Recycler to allow the slip-stacking to occur there, thus reducing the cycle time of the Main Injector to only the 1.3s needed to accelerate the beam and ramp back down [1]. Along with other smaller changes, this allows us to reach 700 kW with only about a 10% increase in intensity per pulse. The biggest challenge is maintaining high efficiency and low beam losses.

Slip-stacking in the Recycler was commissioned in steps, first operating with 6 non-slip-stacked bunches, and then slip-stacking in an additional 2, 4, and 6 bunches successively. At each step, the beam intensity was increased over time while tuning to decrease losses. The increase in beam power and integrated number of protons delivered to NuMI over the past two years are shown in Fig. 2. (Note that the SY120 slow-spill program takes 10% of the timeline, which reduces the beam power to NuMI accordingly. Thus the goal of 700 kW will mean operating at 630 kW when SY120 is running.) The record beam power delivered to the NuMI target averaged over an hour is 613 kW. 700 kW has also been demonstrated for less than an hour. A collimator system is currently being installed in the Recycler to capture beam losses from slip-stacking which will allow us to run routinely at 700 kW in 2017.

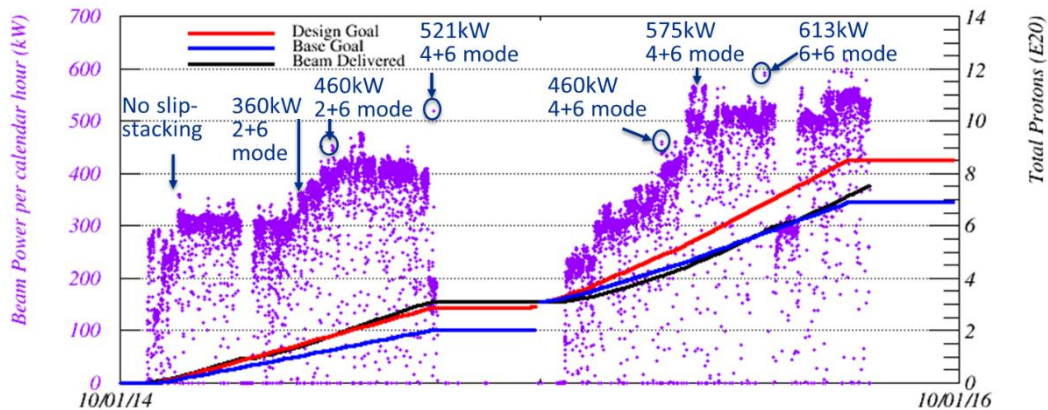

Figure 2: Beam power per hour to the NuMI target (purple) and integrated protons on target (black) compared to base (blue) and design (red) goals since October 2015.

### 3. Proton Improvement Plan

The Proton Improvement Plan (PIP) was initiated in 2011 as a multi-year campaign funded through operations to increase the beam repetition rate from ~7 Hz to 15 Hz and double the proton source flux to more than $2 \times 10^{17}$ protons per hour without increasing total beam losses. Other goals are to eliminate major reliability vulnerabilities and obsolescence issues in order to ensure a useful operating life of the proton source through at least 2030.

The capability to run beam at 15 Hz was achieved in June 2015 after extensive refurbishing of RF cavities and tuners. New anode power supplies are in place to support an increased number of RF stations; a total of 22 cavities will be in place in 2017, compared to 19 prior to PIP. In 2016, the proton source reached its highest proton flux to date of $1.8 \times 10^{17}$ proton per hour (pph). A new shielding assessment which takes into account the use of Total Loss Monitors to restrict beam losses to safe levels is under review and is expected to increase the flux limit to $2.5 \times 10^{17}$ pph. This will enable us to supply beam to NuMI (at 700 kW), BNB, and the future Muon Campus.

Beam physics improvements have been essential in reducing losses. A new cogging system in the Booster was implemented which allowed moving the creation of a notch in the beam where the extraction kicker fires from 700 MeV to 400 MeV, reducing beam energy loss at the notch by 40% and the total cycle loss by 15%.

A laser notching system is being commissioned which will move the process to 750 keV in the Linac medium energy beam transport, thus reducing beam energy loss significantly and removing most of the loss from the Booster. The laser from a burst mode fiber/solid state MOPA laser system is transported to an optical zig-zag cavity (to reduce the required laser energy by a factor of 20) built into the vacuum flange of the RFQ, where it neutralizes the H- beam. Partial notching was successfully demonstrated just prior to the accelerator summer shutdown. Although the laser system was not yet at design intensity (0.5 mJ/pulse), ~70% neutralization in the Linac pulse train was achieved, with the gap surviving at the 40% level in the Booster. The goal is to make full notching operational in 2017. The successful demonstration of the ability to neutralize multiple sections of a 200 MHz H- ion bunch structure out of the RFQ at the Booster revolution period is an important new accelerator development.

### 4. Booster Neutrino Beam

Protons at 8 GeV are transported to the short-baseline neutrino experiments through the Booster Neutrino Beam (BNB). Performance over the last year exceeded expectations due to the quick ramp up of flux, both in terms of 15 Hz repetition rate and intensity, as shown in Fig. 3. A future upgrade is planned for the horn which focuses secondary beam from the BNB target and its associated power supply. The new horn will be longer with inner conductor shape optimized for its length and current for efficient focusing. The horn power supply will be upgraded to allow running at higher current and repetition rate. These improvements are projected to increase neutrino yield by up to 70%.

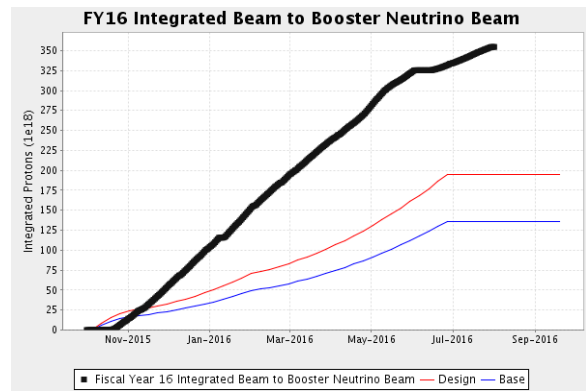

Figure 3: Integrated protons on target (black) to BNB compared to base (blue) and design (red) goals since October 2015.

### 5. Muon Campus

The Muon Campus beamlines to deliver beam to the Mu2e and Muon g-2 experiments are being converted out of the former Antiproton Source ("Pbar"). Since 8 GeV protons will be sent to the Muon Campus from the Recycler instead of 120 GeV protons from the Main Injector, a new extraction line from the Recycler has been constructed. New 2.5 MHz RF cavities are also being installed in the Recycler which are designed to re-bunch the $4 \times 10^{12}$ protons from the Booster into bunches of $10^{12}$ protons with the proper time structure for g-2 and Mu2e. Sixteen of these bunches will be sent to g-2 every 1.4s for an average of 11.4 Hz, but in bursts of 8 bunches at 100 Hz. Although the Pbar target and lithium lens are being reused for g-2, new pulsed power supplies for the lens and momentum-selection magnet are being built in order to meet this repetition rate. Beam transport lines from the target, redesigned to capture as many 3.1 GeV secondary pions and tertiary muons as possible for g-2 as well as to transport 8 GeV primary protons for Mu2e, have been installed. For g-2, the beam will go around the former Pbar Debuncher, renamed the Delivery Ring, four times to allow the secondary protons to separate in time from the muons; the protons are then kicked into an abort as the muons are extracted towards the g-2 muon storage ring. For Mu2e, the primary protons will be resonantly extracted from the Delivery Ring and transported to a target inside their solenoid.

All beamline construction for g-2 is expected to be complete in the spring of 2017, and first beam delivered to the muon storage ring prior to the 2017 summer shutdown.

**6.     Summary**

The Fermilab accelerator complex is performing well. Following Recycler collimator installation during the current shutdown, 700 kW beam power to the NuMI target for NOvA will be operational. Many of the PIP improvements in beam delivery have been realized already; work to ensure reliability remains. This can be seen in the flux to the short baseline neutrino experiments in BNB, which was nearly twice what we projected for this year. Optimizations of an upgraded horn and power supply could increase BNB neutrino yield by up to 70%. Muon Campus construction is on track and will be ready to commission and operate beam to g-2 starting in 2017.